\documentclass{article}
\usepackage{graphicx}
\usepackage{amsmath, amssymb, amsthm}
\usepackage{booktabs, longtable, multirow}
\usepackage{natbib} 
\graphicspath{{figures/}}

\newtheorem{assumption}{Assumption}
\newtheorem{assertion}{Assertion}
\newtheorem{lemma}{Lemma}

\title{Uncovering Patterns of Participant-Invariant Influence in Networks}
\author{Min Shaojie \\
        \small Chongqing University \\
        \small \texttt{alexmin@cqu.edu.cn}
}

\begin{document}

\maketitle

\begin{abstract}
   In this paper, we explore the nature of influence in a network. The concept of participant-invariant influence is derived from an influence matrix $M$ specifically designed to explore this phenomenon. Through nonnegative matrix factorization approximation, we managed to extract a participant-invariant matrix $H$ representing a shared pattern that all participants must obey. The acquired $H$ is highly field-related and can be further utilized to cluster factual networks. Our discovery of the unveiled participant-independent influence within network dynamics opens up new avenues for further research on network behavior and its implications.
\end{abstract}

\section{Introduction}

In the ever-evolving field of network analysis, the concept of influence stands as a cornerstone in understanding the dynamics within various types of networks. This paper delves into the exploration of participant-invariant influence in networks, a concept that transcends the traditional view of influence being tied to specific nodes or participants. By employing an influence matrix specifically designed for this study, we have successfully extracted a participant-invariant matrix. This matrix encapsulates a shared pattern of influence that all network participants adhere to, regardless of their individual roles or positions within the network.

Our approach extends beyond merely measuring influence dynamics; it paves the way for a novel understanding of network behavior. Based on the uncovered shared pattern of influence, this paper also introduces a straightforward method to classify factual networks. The implications of uncovering participant-independent influence within network dynamics are profound, opening new avenues for research into network behavior and its broader implications.

\section{Literature Review}
In the realm of network analysis, understanding and quantifying influence has traditionally centered around various centrality measures. These include not only degree centrality but also betweenness, closeness, and eigenvector centrality \cite{Freeman1978, Newman2010}. These metrics, while foundational, offer primarily a static snapshot of influence within the network.

Transitioning from static measures, research in network analysis has broadened to delve into influence dynamics, exploring how influence propagates and evolves within networks. Studies in this domain, such as those employing perturbative formalism \cite{Barabasi2016}, optimal percolation models \cite{MoroneMakse2015}, patterns of information flow \cite{Gao2017}, and the dynamics of opinions in echo chambers and polarization \cite{Baumann2020}, reveal intricate patterns of influence spread. These studies highlight the transient nature of network interactions. While these models offer deeper insights into the propagation and evolution of influence within networks, they primarily concentrate on the influence exerted by specific nodes and their consequent applications such as finding the most influential nodes \cite{Zheng2019} or influence maximization in terms of information propagation \cite{Kalimeris2019}.

Existing works on influence dynamics essentially consider the influence of participants over others, rather than the pattern of influence dynamics across the network. This realization guides our exploration towards an approach that aims to encapsulate the pattern of influence increment. Furthermore, our research goes beyond the measurement of influence dynamics by successfully extracting influence patterns shared across the network. This observation fills the gap in understanding the invariant characteristics of influence that persist across all participants in the network, complementing and extending the insights provided by previous dynamic models.

In the context of graph embedding, traditional approaches, as outlined by Cui et al. \cite{Cui2018}, have been instrumental in translating complex network structures into lower-dimensional vector spaces. Our methodology aligns with this principle as extracting shared influence patterns provides a lower-dimensional representation of high-dimensional network data. Consequently, our approach contributes a fresh perspective to comprehending network dynamics, seeking to bridge the gap between examining static structural analysis and the dynamic nature of influence.

\section{Methodology}
Given a system of temporal network, denoted as \( G_t = (V_t, E_t), \) in which \( G_t \) represents the network's transient state at time \( t \), we design a matrix \( M_t \in \mathbb{R}^{n\times t_{max}} \) to quantify participants' edge-formation dynamics that:
\[ M_{i,t} = \frac{dev(i,\ t) - dev(i,\ t-1)}{n_i} \]
where \( n_i \) is the number of nodes at the time when \( i \) appears in the network and \( dev(i,\ t) \) is the degree of \( i \) at time \( t \).

\( M_t \) can be acquired from \( M_{i,t} = \Pi_t(i,i) \) in which \( \Pi \) is the diagonal matrix defined as:
\[ \Pi_t =\begin{cases} 
\frac{D_t - D_{t-1}}{n_i} &,\ t>0 \\
0 &,\ t=0 
\end{cases} \]
where \( D_t \in \mathbb{R}^{n \times n} \) is the degree matrix of \( G_t \).

In addition, we can further align \( M \) to \( M^* \) such that each participant's \( M_{i,} \) is aligned according to the time of its first occurrence in the system, \( t_0(i) \).
\[ M_{i, j}^* =\begin{cases}
M_{i,\ j+t_0(i)} &,\  j\le t_{max}-t_0(i) \\
0 &,\ t_{max}-t_0(i)<j\le t_{max} 
\end{cases} \]

\begin{assumption}\label{assump:h}
Given a system of a temporal network, \( G_t = (V_t, E_t), \) the system comes with an inherent non-negative function, denoted as \( \varphi_{\mathcal{G}} (\cdot) \), that is participant-invariant and a participant vector \( W \in \mathbb{R}^{n \times 1} \) such that:
\[ 
\begin{aligned}
WH &= M^* \\
\text{s.t. } H_i &= \varphi_{\mathcal{G}}(i),\ H \in \mathbb{R}^{1\times t_{max}}
\end{aligned}
\]
\end{assumption}

Consequently, acquiring the participant-invariant \( \varphi_{\mathcal{G}} (\cdot) \) effectively translates to the extraction of \( H \) from \( M^* \). Deriving \( (W, H) \) directly from \( M^* \) is a non-trivial endeavor, demanding the imposition of additional constraints on the network under investigation. To address this, we employ Nonnegative Matrix Factorization (NMF) to obtain a pair of \( (W, H) \) to approximate the non-negative matrix \( M^* \), so that \( H \) can be acquired as minimizing:
\[ \min _{\mathbf{W}, \mathbf{H} \in \mathbb{R}^{1\times t_{max}}}\left\|\mathbf{M}^*- \mathbf{WH}\right\|_{F}^{2} \]

Due to the nature of NMF, it is impossible to ask for the uniqueness of H. However, our assumption of participant-invariant makes deduction theoretically's uniqueness unnecessary.

\begin{assertion}\label{assert:unique}
Uniqueness of \( H \)

Due to the participant-invariant nature of \( \varphi_{\mathcal{G}} (\cdot) \), hence \( H \), any \( G_t \)'s sub-system which consists of partial participants can also derive an \( H^{'} \)' that is identical to \( H \). In other words, for any \( M \)'s subset, \( \mathcal{M} \in \mathbb{R}^{s\times t_{max}},\ s \le n \), we all have:
\[ 
\begin{aligned} \label{eq:unique}
\mathbf{H}_{\mathcal{M}} &\equiv \mathbf{H} \\
\text{s.t. } \mathbf{H}_{\mathcal{M}} &\equiv \text{argmin}_{\mathbf{H} \in \mathbb{R}^{1\times t_{max}}}\left\|\mathcal{M}^*- \mathcal{W}\mathbf{H}_{\mathcal{M}}\right\|_{F}^{2}
\end{aligned}
\]
\end{assertion}

If we randomly subset participants for \( \rho \) times, we will have a sequence of \( \mathbf{M} \) subsets, \( \{ \mathcal{M_1}, \mathcal{M_2},\cdots ,\ \mathcal{M_k} \} \), from which we can derive a sequence of \( H \), \( \{ \mathbf{H}_{\mathcal{M_1}}, \mathbf{H}_{\mathcal{M_2}},\cdots ,\ \mathbf{H}_{\mathcal{M_\rho}} \} \), that satisfies:
\[ \frac{1}{\rho} \sum_{i}^{\rho}\left\| \mathbf{H}_{\mathcal{M_i}}-\mathbf{H} \right\| < \epsilon \]
in which, \( \epsilon \) is a reasonably small number and \( \left\| \cdot \right\| \) represents a distance measurement for \( H \).

\section{Experiments}
\subsection{Experiment Settings}

\begin{table}[h]
   \centering
   \caption{Network Statistics}
   \label{tab:networkstats}
   \begin{tabular}{@{}llrr@{}}
   \toprule
   \textbf{Network} & \textbf{Category} & \textbf{Nodes} & \textbf{Temporal Edges} \\ 
   \midrule
   sx-mathoverflow & Academic Network & 24818 & 506550 \\
   sx-mathoverflow-c2q &  & 16836 & 203639 \\
   sx-mathoverflow-c2a &  & 13840 & 195330 \\
   sx-mathoverflow-a2q &  & 21688 & 107581 \\
   sx-askubuntu &  & 159316 & 964437 \\
   sx-askubuntu-c2q &  & 79155 & 327513 \\
   sx-askubuntu-c2a &  & 75555 & 356822 \\
   sx-askubuntu-a2q &  & 137517 & 280102 \\
   sx-superuser &  & 194085 & 1443339 \\
   sx-superuser-c2q &  & 94548 & 479067 \\
   sx-superuser-c2a &  & 101052 & 534239 \\
   sx-superuser-a2q &  & 167981 & 430033 \\
   CollegeMsg & Social Network & 1899 & 59835 \\
   soc-sign-bitcoin-alpha &  & 3783 & 24186 \\
   soc-sign-bitcoin-otc &  & 5881 & 35592 \\
   ia-chess-combined &  & 7301 & 65053 \\
   ia-retweet-pol &  & 18470 & 61157 \\
   cit-HepPh & Citation Network & 30566 & 347414 \\
   cit-HepTh &  & 18477 & 136427 \\
   cit-Patents-DK &  & 28873 & 29988 \\
   cit-Patents-ES &  & 17998 & 17027 \\
   cit-Patents-IL &  & 46986 & 55935 \\
   cit-Patents-SU &  & 21719 & 19964 \\
   email-Eu-core-temporal & Email Network & 986 & 332334 \\
   email-Eu-Dept1 &  & 309 & 61046 \\
   email-Eu-Dept2 &  & 162 & 46772 \\
   email-Eu-Dept3 &  & 89 & 12216 \\
   email-Eu-Dept4 &  & 142 & 48141 \\
   \bottomrule
   \end{tabular}
   \end{table}

In this chapter, we present a comprehensive set of experiments conducted to validate the effectiveness of our methodology in extracting the participant-invariant function \( H \) from various networks. These 28 networks are categorized into Academic Networks, Social Networks, Citation Networks, and Email Networks, as detailed in the table below. The sizes of these networks vary, with nodes ranging from as few as 89 in the smallest network, email-Eu-Dept3, to over 194,000 in the largest, sx-superuser. The number of edges also shows significant variation, from 12,216 in smaller networks to over 1.4 million in the larger ones (refer to Table \ref{tab:networkstats} for detailed network statistics).

To ensure objectivity and uniformity in our analysis, a consistent preprocessing approach was applied across all networks. This involved:
\begin{itemize}
    \item Edges with missing temporal information are removed;
    \item Direct edges are treated as undirected ones;
    \item Parallel edges are preserved where they existed;
    \item To maintain the same length of \( H \) across all networks, 400 snapshots are taken for each network with the same number of edges in each interval.
\end{itemize}

These preprocessing steps were crucial in maintaining the integrity and comparability of our results across different network types and sizes.

\subsection{Participant-Invariant \( H \) in Networks}

\begin{figure}[ht]
\centering
\includegraphics[width=0.8\textwidth]{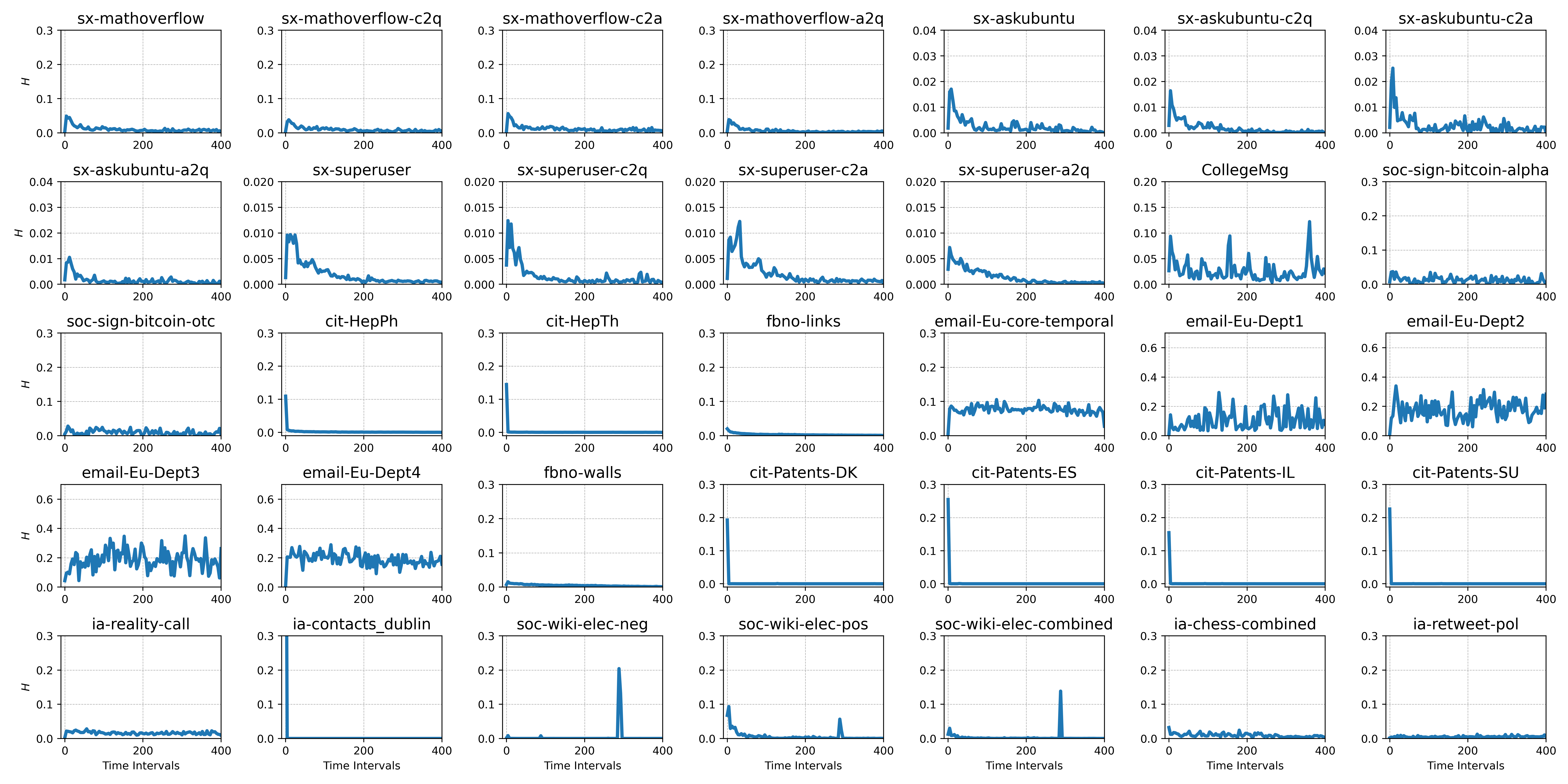}
\caption{$H$ extracted from 28 networks.}
\label{fig:all_h}
\end{figure}

Within each of the 28 different networks, the participant-invariant function \( H \) was successfully extracted with a tolerance rate below 0.01\% under Frobenius form (as illustrated in Figure \ref{fig:all_h}), demonstrating the versatility and adaptability of our approach for extracting participant-invariant influence patterns across different network types and sizes.

\subsubsection{Uniqueness of \( H \)}

Here, we experimentally test the uniqueness of each \( H \) with Assertion \ref{assert:unique}. Although \( H \) is essentially a time-series vector, to test the uniqueness of \( H \) for each network, we should not consider the temporal relation among elements within the vector because we intend the point-wise consistency across all \( H_{\mathcal{M_i}} \) (from the network subset \( \mathcal{M_i} \)) and \( H \). Thus, to assert the uniqueness of \( H \), we employ the concept of cross-validation where we randomly select \( 1 - \frac{1}{\rho} \) of network participants for \( \rho \) times to get a sequence of \( \{ \mathbf{H}_{\mathcal{M_1}}, \mathbf{H}_{\mathcal{M_2}}, \cdots , \mathbf{H}_{\mathcal{M_\rho}} \} \). In practice, we set \( \rho \) to 5, 10, and 20 to ensure the influence patterns are preserved in any sub-network.

Since L-p norms can be sensitive to the dimensionality of the data, the value of norms will inevitably increase as more snapshots are taken for each network. Thus, we utilize normalized L1-norm, normalized L2-norm, and cosine similarity as distance measurements of \( H \)'s uniqueness to account for the potential impact of this cumulative effect. The normalized norm is calculated as:

\[ \frac{1}{\rho} \sum_{i}^{\rho} \frac{\left\| \mathbf{H}_{\mathcal{M_i}}-\mathbf{H} \right\| _{L_p}}{t} \]

and table \ref{tab:unique} shows the results of distance-measurement criteria in Formula \ref{tab:unique} for all 28 networks.

{\scriptsize
\centering
\begin{longtable}{@{}lrrrrrrrrr@{}}
\caption{Uniqueness of \( H \) Across Networks}\label{tab:unique} \\
\toprule
Measurement & &  & l1 norm  &  &  & l2 norm  & &  & cosine  \\
$\rho$ & 5 & 10 & 20 & 5 & 10 & 20 & 5 & 10 & 20 \\ \midrule
\endfirsthead
\multicolumn{10}{c}%
{{\bfseries Table \thetable\ continued from previous page}} \\
\toprule
Measurement & &  & l1 norm  &  &  & l2 norm  & &  & cosine  \\
$\rho$ & 5 & 10 & 20 & 5 & 10 & 20 & 5 & 10 & 20 \\ \midrule
\endhead
\bottomrule
\endfoot
\endlastfoot
sx-superuser & 4.3e-4 & 4.6e-4 & 4.4e-4 & 4.4e-5 & 4.9e-5 & 4.5e-5 & 9.7e-1 & 9.2e-1 & 9.6e-1  \\
-c2q \\

sx-askubuntu & 7.4e-4 & 7.5e-4 & 6.6e-4 & 5.9e-5 & 7.1e-5 & 6.2e-5 & 9.6e-1 & 9.3e-1 & 9.5e-1  \\
-c2q \\

email-Eu-Dept4 & 4.1e-2 & 4.4e-2 & 4.1e-2 & 3.7e-3 & 2.7e-3 & 2.4e-3 & 9.7e-1 & 9.6e-1 & 9.7e-1  \\

cit-Patents-IL & 3.5e-5 & 2.9e-5 & 2.7e-5 & 8.7e-6 & 8.0e-6 & 7.5e-6 & 1.0e+0 & 1.0e+0 & 1.0e+0  \\

email-Eu-Dept3 & 6.9e-2 & 5.9e-2 & 6.9e-2 & 5.6e-3 & 4.6e-3 & 4.5e-3 & 8.6e-1 & 9.2e-1 & 9.1e-1  \\

email-Eu-Dept2 & 3.2e-2 & 3.1e-2 & 3.2e-2 & 2.0e-3 & 2.7e-3 & 2.3e-3 & 9.4e-1 & 9.8e-1 & 9.7e-1  \\

cit-Patents-ES & 2.7e-5 & 3.1e-5 & 3.1e-5 & 1.4e-5 & 1.4e-5 & 1.6e-5 & 1.0e+0 & 1.0e+0 & 1.0e+0  \\

cit-Patents-SU & 3.9e-5 & 3.6e-5 & 3.3e-5 & 1.2e-5 & 1.1e-5 & 1.4e-5 & 1.0e+0 & 1.0e+0 & 1.0e+0  \\

sx-mathoverflow & 3.0e-3 & 3.3e-3 & 3.3e-3 & 3.7e-4 & 2.6e-4 & 2.1e-4 & 9.4e-1 & 9.6e-1 & 9.5e-1  \\
-c21 \\

fbno-walls & 3.8e-4 & 4.2e-4 & 3.7e-4 & 3.1e-5 & 2.7e-5 & 2.6e-5 & 1.0e+0 & 1.0e+0 & 1.0e+0  \\

sx-askubuntu & 5.7e-4 & 8.1e-4 & 6.5e-4 & 4.8e-5 & 7.6e-5 & 6.0e-5 & 9.4e-1 & 9.5e-1 & 9.3e-1  \\

sx-mathoverflow & 1.7e-3 & 1.3e-3 & 1.7e-3 & 1.1e-4 & 1.0e-4 & 1.2e-4 & 9.5e-1 & 9.6e-1 & 9.6e-1  \\
-a2q \\

soc-sign-bitcoin-otc & 7.3e-3 & 5.1e-3 & 6.1e-3 & 4.1e-4 & 2.4e-4 & 4.6e-4 & 8.8e-1 & 8.2e-1 & 7.8e-1  \\

cit-Patents-DK & 3.8e-5 & 3.0e-5 & 4.0e-5 & 1.4e-5 & 1.9e-5 & 1.2e-5 & 1.0e+0 & 1.0e+0 & 1.0e+0  \\

cit-HepTh & 8.7e-5 & 9.3e-5 & 1.0e-4 & 3.2e-5 & 3.4e-5 & 4.5e-5 & 1.0e+0 & 1.0e+0 & 1.0e+0  \\

sx-superuser & 5.4e-4 & 4.1e-4 & 4.2e-4 & 3.7e-5 & 3.6e-5 & 4.2e-5 & 9.3e-1 & 9.3e-1 & 9.4e-1  \\
-a2q \\

sx-askubuntu & 4.0e-4 & 4.0e-4 & 4.8e-4 & 5.3e-5 & 4.4e-5 & 4.0e-5 & 8.9e-1 & 9.5e-1 & 9.5e-1  \\
-a2q \\

sx-superuser & 1.1e-3 & 8.1e-4 & 8.6e-4 & 9.3e-5 & 8.7e-5 & 7.1e-5 & 8.5e-1 & 9.4e-1 & 8.6e-1  \\
-c2a \\

sx-askubuntu & 1.1e-3 & 1.5e-3 & 9.2e-4 & 4.9e-5 & 5.0e-5 & 9.1e-5 & 8.7e-1 & 9.9e-1 & 9.5e-1  \\
-c2a \\

email-Eu-Dept1 & 3.7e-2 & 3.5e-2 & 3.5e-2 & 2.0e-3 & 2.7e-3 & 2.0e-3 & 9.5e-1 & 8.7e-1 & 8.9e-1  \\

email-Eu-core & 1.1e-2 & 8.7e-3 & 8.9e-3 & 5.9e-4 & 5.8e-4 & 5.7e-4 & 9.9e-1 & 9.9e-1 & 9.9e-1  \\
-temporal \\

fbno-links & 1.8e-4 & 1.8e-4 & 1.8e-4 & 1.2e-5 & 1.3e-5 & 1.2e-5 & 1.0e+0 & 1.0e+0 & 1.0e+0  \\

sx-superuser & 7.9e-4 & 6.3e-4 & 5.1e-4 & 4.4e-5 & 5.3e-5 & 5.4e-5 & 9.5e-1 & 9.5e-1 & 9.5e-1  \\

CollegeMsg & 1.3e-2 & 1.2e-2 & 1.1e-2 & 9.9e-4 & 1.0e-3 & 9.5e-4 & 8.0e-1 & 8.7e-1 & 8.6e-1  \\

sx-mathoverflow & 2.4e-3 & 2.4e-3 & 2.2e-3 & 1.8e-4 & 1.5e-4 & 1.4e-4 & 9.7e-1 & 9.6e-1 & 9.7e-1  \\
-c2q \\

soc-sign-bitcoin & 9.8e-3 & 8.5e-3 & 7.8e-3 & 6.9e-4 & 5.5e-4 & 5.6e-4 & 6.9e-1 & 7.9e-1 & 8.3e-1  \\
-alpha \\

cit-HepPh & 2.6e-4 & 2.1e-4 & 1.9e-4 & 1.9e-5 & 3.7e-5 & 2.2e-5 & 1.0e+0 & 1.0e+0 & 1.0e+0  \\

sx-mathoverflow & 2.6e-3 & 2.4e-3 & 2.3e-3 & 2.5e-4 & 1.8e-4 & 1.7e-4 & 9.7e-1 & 9.8e-1 & 9.7e-1  \\ \bottomrule
\end{longtable}
}

\subsection{Domain-Specific Influence Patterns}

\begin{figure}[ht]
\centering
\includegraphics[width=0.8\textwidth]{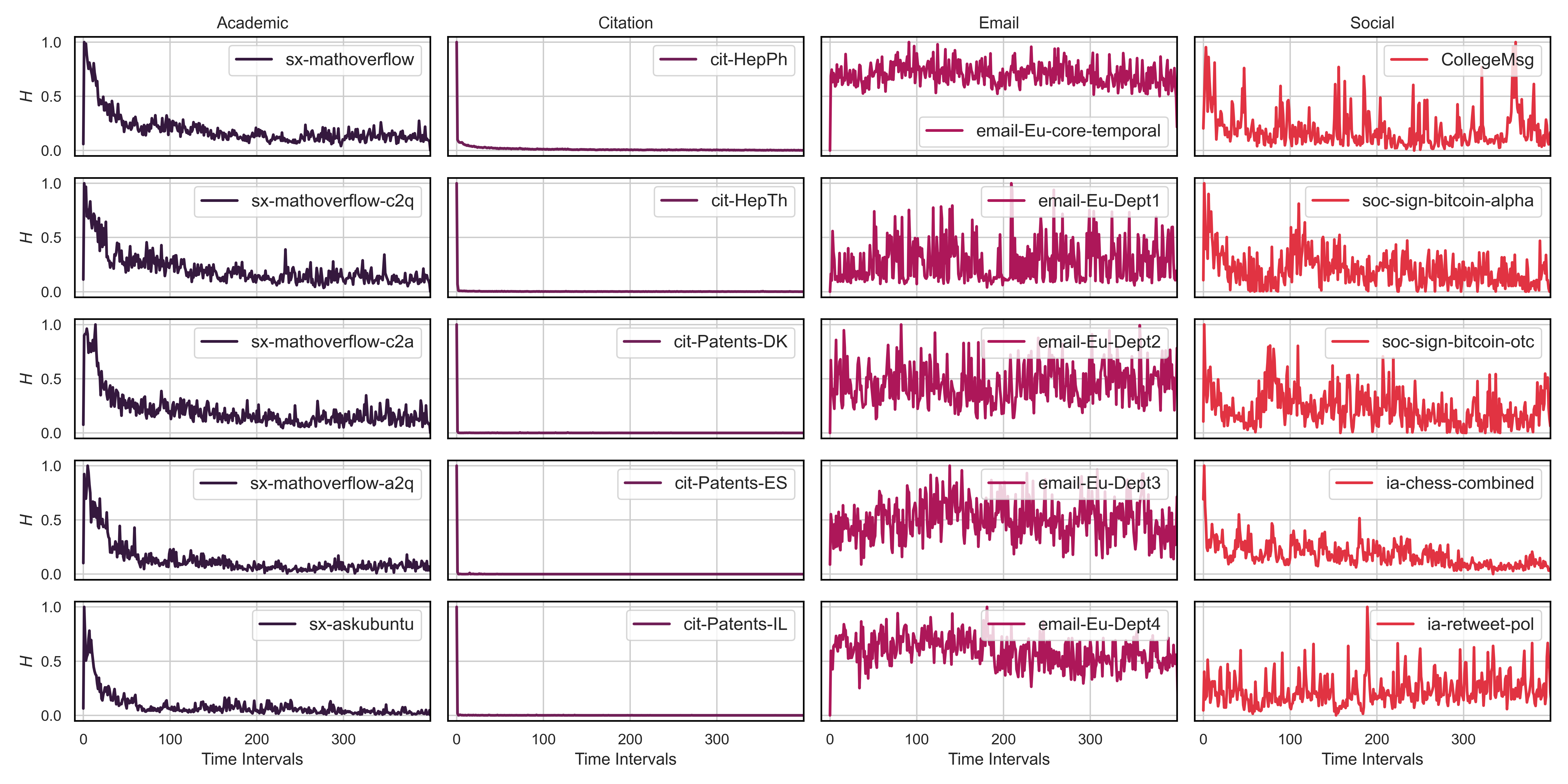}
\caption{$H$ is highly domain-related.}
\label{fig:h_shared}
\end{figure}

Besides unique patterns of influence revealed by \( H \) for each network, the characteristics of \( H \) are closely related to the domain of networks. To measure the similarity of different \( H \), we intend to preserve the patterns while minimizing the impact of the out-of-phrase circumstance owing to the varied time spans of different networks (one \( H \) might be compressed or stretched relative to the other). To attend to this matter, we employ Dynamic Time Warping (DTW).

DTW is an algorithm used to measure the similarity between two temporal sequences, which may vary in time span. It aligns two sequences by stretching or compressing them in the time dimension. The goal is to find an optimal match between these sequences, where the 'cost' of aligning each point in one sequence with a point in the other is minimized. In the context of analyzing the characteristics of \( H \), DTW can be particularly useful for comparing the influence patterns extracted from different networks, especially when these patterns may have similar shapes but are not aligned in the time axis.

As DTW is sensitive to values of each data point in the time-series while we only want to quantify the patterns' similarities, (e.g. the DTW distance between \(\sin x\) and \(\sin x + 2\) is always much larger than the distance between \(\sin x\) and \(\sin x + 1\), however, for our concerns the difference is trivial) we normalized each \( H \) to (0, 1) before calculation.

\begin{figure}[ht]
\centering
\includegraphics[width=0.8\textwidth]{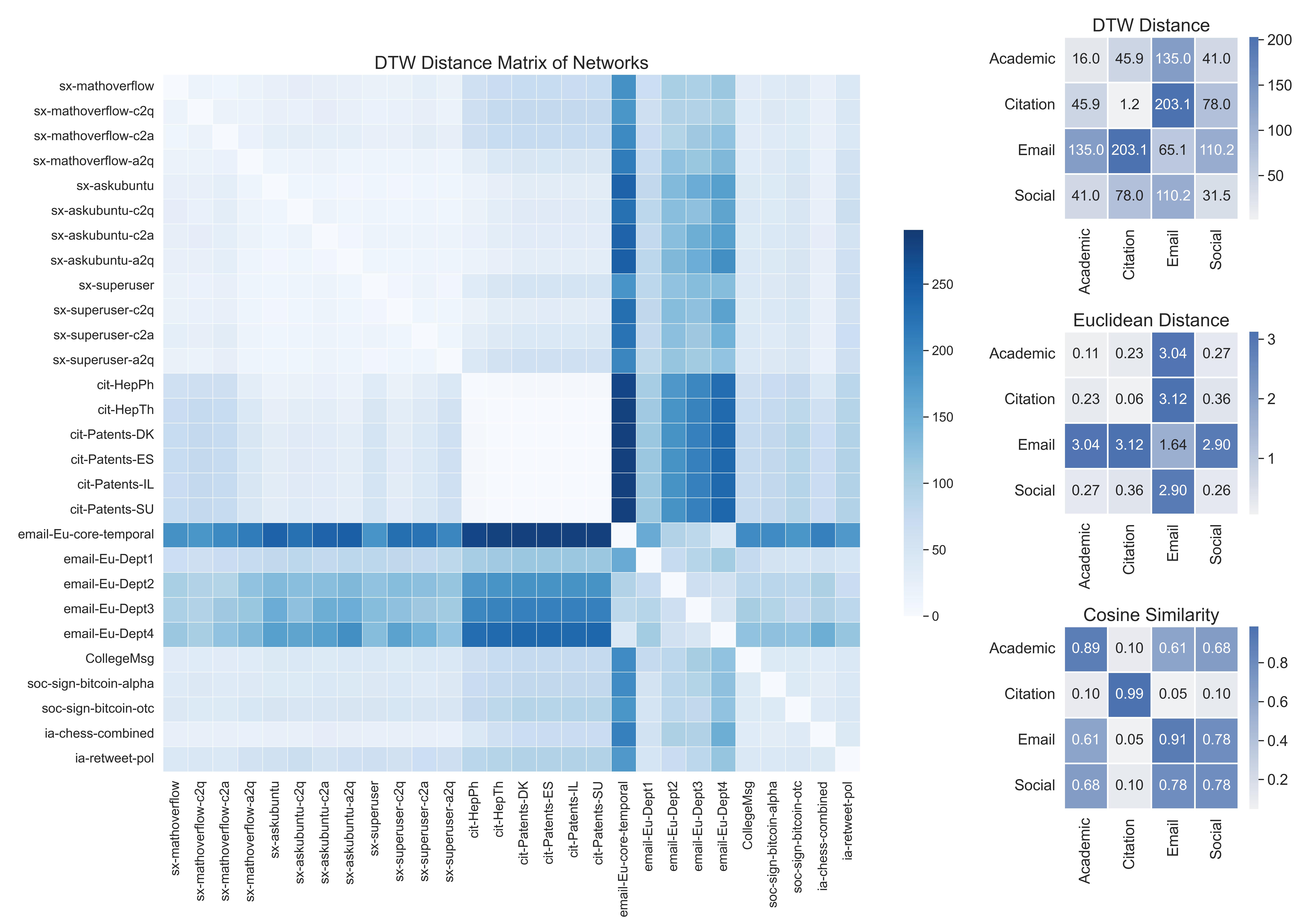}
\caption{Similarly measurements for all H.}
\label{fig:h_similarities}
\end{figure}

The analysis revealed that the characteristics of \( H \) are closely related to the network domain. Under the same field, networks exhibit lower DTW values, hence more similar patterns of \( H \). For example, in academic networks like `cit-HepPh` and `cit-HepTh`, \( H \) captured the influence dynamics specific to citation practices where the influence rapidly decays after it is published, while in social networks like `CollegeMsg`, it reflects more on the social interaction patterns where the influence of a participant lasts longer. We also test other category-level measurements as complements to DTW (i.e., cosine similarity and Euclidean distance). These alternative measurements provided similar results, further validating the consistency of the findings. This field-related nature of \( H \) underscores its capability to adapt to the inherent dynamics of different network categories and can be utilized to predict the field of an unknown network.

\section{Discussion}

\subsection{Methodologies Extensions}

\subsubsection{Extending the Dynamic Measuring Matrix}

In our initial framework, the dynamic measuring matrix \( M \) was primarily focused on capturing influence dynamics via the degree of a node at time \( t \). While this approach provides easy-to-follow insights, it is essential to encompass a broader spectrum of influence measures to fully understand network dynamics.

We propose to extend the functional scope of the matrix \( M \) by incorporating a more comprehensive set of influence metrics. Instead of solely focusing on node degree, \( M \) is now defined to incorporate a generic influence measurement symbolized by \( \delta(i, t) \) such as betweenness centrality, closeness centrality, or other network-specific influence metrics, such that:
\[ M_{i,t} = \frac{\delta(i, t) - \delta(i, t-1)}{n_i} \]
This extension allows matrix \( M \) to adapt to various network structures and influence dynamics, offering a more comprehensive understanding of the influence dynamics within the network.

\subsubsection{Decomposition in Multi-Role Networks}

In terms of multi-role networks where different roles of participants inevitably exhibit heterogeneous patterns of influence dynamics, we introduce the concept of extracting multiple participant-invariant functions, \( \varphi_{\mathcal{G}, k} (\cdot) \), in conjunction with a multi-dimensional participant vector \( W \in \mathbb{R}^{n \times k} \) to address this issue, in which each row's \( m \) element, \( m \in \{1, 2, \cdots, k\} \) represents how much the corresponding participant is related to the role \( m \).

\begin{lemma}
Given a system of a k-role temporal network, \( G_t = (V_t, E_t), \) the same NMF approach can be applied to extract an inherent non-negative function set, denoted as \( \varphi_{\mathcal{G}, k} (\cdot) \), that is role-related but participant-invariant and a participant vector \( W \in \mathbb{R}^{n \times k} \) such that:
\[ 
\begin{aligned}
WH &= M^* \\
\text{s.t. } H_{m, i} &= \varphi_{\mathcal{G}, m}(i),\ H \in \mathbb{R}^{k\times t_{max}},\ m \in \{1, 2, \cdots, k\}
\end{aligned}
\]
\end{lemma}

This approach allows for the extraction of distinct influence patterns corresponding to the various roles within a network. By doing so, it enhances the ability to discern and analyze participant-invariant influence patterns, contributing to a deeper understanding of the dynamics in multi-role networks.

\subsection{Averaged DTW as Measurement}

For DTW, we consider two sequences or vectors \( v_1 \) in \( \mathbb{R}^m \) and \( v_2 \) in \( \mathbb{R}^n \). As the basic idea of DTW is to find an alignment between these two sequences that minimizes the total distance between them, typically we construct a matrix where the element at position (i, j) represents the distance between \( v_1[i] \) and \( v_2[j] \) and find a path through this matrix that minimizes the cumulative distance. This path represents the best alignment between the two sequences.

\begin{figure}[ht]
\centering
\includegraphics[width=0.8\textwidth]{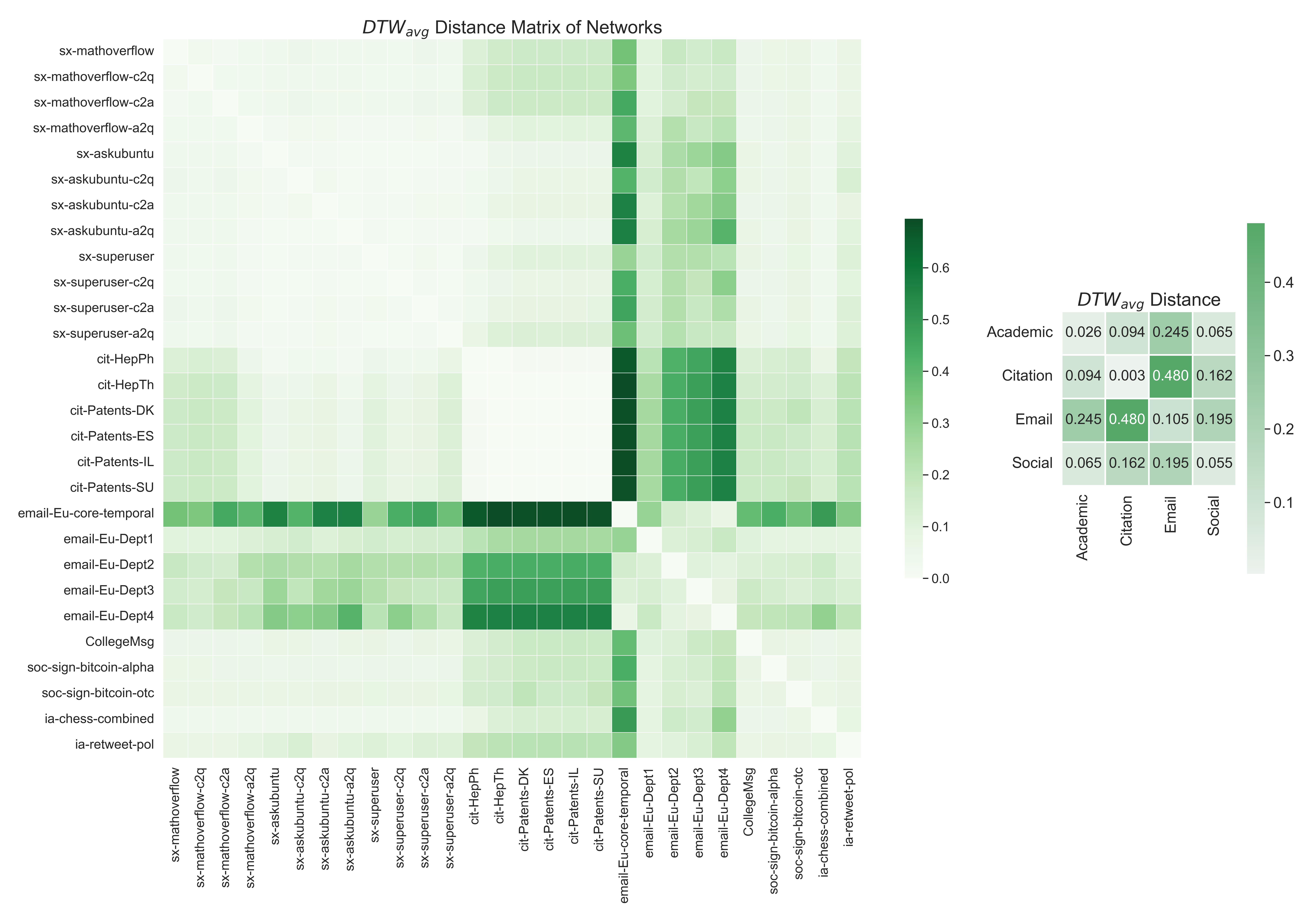}
\caption{Averaged DTW as Measurement}
\label{fig:dtw_avg}
\end{figure}

The total number of steps (and thus the number of point-wise distances summed) is not fixed just by the lengths of the input sequences. It depends on the specific path the DTW algorithm finds, which will have more than max(m, n) steps and fewer than m + n steps. As a consequence, the cumulative distance may be the result of 400 to 800 point-wise distances summed even if a uniformed 400 snapshots are taken for each network. To attend to this matter, we also calculated the average DTW distance for each \( H \) (i.e. \( DTW_{avg} = \frac{DTW}{\text{\# of steps}} \)) as shown in figure \ref{fig:dtw_avg}. The result also supports our claim about domain-specific patterns.

\section{Summary}

This paper presented a comprehensive exploration of participant-invariant influence within networks, revealing unique patterns of influence across a diverse range of networks. Our methodology, centered around the participant-invariant function \( H \), demonstrated its versatility and adaptability in extracting influence patterns across various network types and sizes. The uniqueness of \( H \) was validated through rigorous experimentation, showcasing its consistency across different sub-systems within a network.

Furthermore, we explored the domain-specific nature of influence patterns, employing Dynamic Time Warping (DTW) to quantify the similarity of \( H \) across networks from different domains. The results confirmed that networks within the same field exhibit more similar patterns of influence, a finding that underscores the adaptability of \( H \) to the inherent dynamics of different network categories. This insight into domain-specific influence patterns not only enriches our understanding of network behavior but also offers a predictive tool for identifying the field of an unknown network.

Overall, the findings and methodologies presented in this paper contribute significantly to the field of network analysis, providing a fresh perspective on understanding the dynamics of influence and opening up new possibilities for future research.

\bibliographystyle{apsrev4-1}
\bibliography{references}

\end{document}